\documentclass[sigconf]{acmart} 

\usepackage{float, graphicx}
\usepackage{lipsum}
\usepackage{multirow}
\usepackage{tabularx}
\usepackage{csquotes}

\AtBeginDocument{%
  }

\usepackage{bm}
\usepackage{amsmath,scalerel}
\usepackage{xfrac}
\usepackage{bbm}
\usepackage{graphicx}
\usepackage{subfig}

\begin{document}



\title{Conveying Meaning through Gestures: An Investigation into Semantic Co-Speech Gesture Generation}

\author{Hendric Voss}
\email{hvoss@techfak.uni-bielefeld.de}
\affiliation{%
  \institution{Bielefeld University}
  \streetaddress{Universitätsstraße 25}
  \country{Germany}
}

\author{Lisa Michelle Bohnenkamp}
\email{lisa_michelle.bohnenkamp@uni-bielefeld.de}
\affiliation{%
  \institution{GID GeoInformationsDienst GmbH}
  \streetaddress{Götzenbreite 10}
  \country{Germany}
}

\author{Stefan Kopp}
\email{skopp@techfak.uni-bielefeld.de}
\affiliation{%
  \institution{Social Cognitive Systems Group}
  \institution{Bielefeld University}
  \streetaddress{Universitätsstraße 25}
  \country{Germany}
}

\renewcommand{\shortauthors}{Voss et al.}

\begin{abstract}
This study explores two frameworks for co-speech gesture generation, AQ-GT and its semantically-augmented variant AQ-GT-a, to evaluate their ability to convey meaning through gestures and how humans perceive the resulting movements. Using sentences from the SAGA spatial communication corpus, contextually similar sentences, and novel movement-focused sentences, we conducted a user-centered evaluation of concept recognition and human-likeness. Results revealed a nuanced relationship between semantic annotations and performance. The original AQ-GT framework, lacking explicit semantic input, was surprisingly more effective at conveying concepts within its training domain. Conversely, the AQ-GT-a framework demonstrated better generalization, particularly for representing shape and size in novel contexts. While participants rated gestures from AQ-GT-a as more expressive and helpful, they did not perceive them as more human-like. These findings suggest that explicit semantic enrichment does not guarantee improved gesture generation and that its effectiveness is highly dependent on the context, indicating a potential trade-off between specialization and generalization.
\end{abstract}




\maketitle


\section{Introduction}
Human communication is inherently multimodal, involving not only verbal cues but also non-verbal gestures that convey additional information, resolve ambiguity, and emphasise relevance. As robots and virtual agents strive to improve their human-like interaction, the incorporation of gestures has become a crucial aspect of their design to enhance the effectiveness of human-agent communication. The use of gestures in artificial systems is rooted in the understanding that humans use different types of gestures, such as iconic, metaphorical, deictic, and beat gestures \cite{mcneill1990speech}, to convey meaning and context. However, generating synthetic gestures that are both natural and meaningful has proven to be a challenging task.

Early approaches to gesture generation relied on rule-based systems that focused on producing communicative gestures, which resulted in highly flexible and adaptable but sometimes unnatural movements \cite{ruttkay_gnetic_2009}. In contrast, data-driven approaches have successfully generated more human-like and natural gestures, but often lack semantic meaning \cite{nyatsanga_comprehensive_2023}. Recent efforts have sought to combine the strengths of both approaches, such as the use of deep learning methods that incorporate audio and text input \cite{ahuja_style_2020,alexanderson_style-controllable_2020,yoon_speech_2020}.

A notable example is the work of \cite{vos_augmented_2023}, who developed a data-driven deep learning approach that incorporates target form and meaning features to improve the semantic meaning of generated gestures. Their framework uses annotated labels from the SaGA corpus \cite{lucking_bielefeld_2010} to specify target form information such as hand shape, wrist position and movement extent. In addition, they use labels for semantic information, including spoken entity, relative position, and entity occurrence.

While this approach has shown promise in producing more natural and human-like gestures, it remains unclear whether the framework can produce specific gestures that accurately represent complex concepts. Furthermore, evaluation of such frameworks often relies on user studies that assess overall performance, but may not provide insight into the representation of specific concepts.

This paper aims to investigate the ability of synthetic gesture generation frameworks, AQ-GT and AQ-GT-a, to convey meaning through gestures, with a focus on the representation of specific concepts. We will investigate whether human-understandable gestural representations of concepts can be generated and evaluated using a systematic approach. In addition, our study will employ a user-centred evaluation methodology to assess the expressiveness, synchronicity, and humaneness of synthetic gestures in different contexts.

We hypothesize (H1) that the gestures produced by AQGT-a are semantically richer than those produced by AQGT, resulting in more concepts being recognized in videos produced using AQGT-a than AQGT. Specifically:

\begin{itemize}
    \item The gestures produced by AQGT-a with annotations (H1.2) and without (H1.3) are expected to be more semantically rich than those produced by AQGT.
    \item The inclusion of SAGA-Entities in the context (H1.4, H1.5, and H1.6) is predicted to enhance concept recognizability compared to contexts without SAGA-Entities or with movements and actions.
\end{itemize}

The second half of the study investigates how generated gestures are perceived by humans. We hypothesize (H2) that the gestures produced by AQGT-a are more human-like, synchronous, and expressive than those produced by AQGT or other contexts.

\begin{itemize}
    \item The inclusion of annotations in AQGT-a (H2.1 and H2.2) is expected to improve the human-likeness and expressiveness of generated gestures.
    \item The SAGA-context with SAGA-Entities (H2.5) is predicted to be more expressive than the context without Entities, while being similar in human-likeness and synchronization to other contexts (H2.7).
    \item The exclusion of SAGA-Entities from the context (H2.6) is expected to result in less expressive gestures compared to the context with SAGA-Entities.
\end{itemize}

These hypotheses will be tested through various metrics, including concept recognition rates and human-likeness perception ratings using the HEMVIP-Questionnaire.

\section{Related Work} 

The generation of human-like gestures has been a long-standing problem in virtual agent research, with applications in human-agent interaction and multimodal communication. Early approaches relied on rule-based models, such as the Behavior Markup Language (BML) \cite{koppCommonFrameworkMultimodal2006a} and the Behavior Expression Animation Toolkit (BEAT) \cite{noauthor_beat_nodate}, which utilized hand-crafted gesture templates to synthesize gestures for given communicative functions. These approaches allowed for direct adaptation of synthesized gestures to specific speakers, styles, or functions, ensuring communicative efficacy \cite{koppCommonFrameworkMultimodal2006a, noauthor_beat_nodate, neff_evaluating_2010}. Despite their limited naturalness, these approaches were widely used in early commercial products, such as the Pepper and Nao robots, due to their adaptability and low computational requirements \cite{pandey_mass-produced_2018}.

With the ongoing development of deep learning, gesture generation shifted towards data-driven methods, particularly for speech-driven gesture synthesis, where gestures are generated in response to spoken utterances. Early neural approaches, such as the recurrent neural network models used by \citet{ferstlInvestigatingUseRecurrent2018a}, focused on generating short sequences of movements using prosodic features. 
Subsequent work has explored more sophisticated architectures, including invertible 1x1 convolutions \cite{henterMoGlowProbabilisticControllable2020} and variational autoencoders combined with deep reinforcement learning for goal-directed control of gestures \cite{ling_character_2020}. 
Recent advances have demonstrated the effectiveness of general adversarial networks (GANs) in producing highly realistic gestures by integrating multiple modalities such as audio, text and speaker identity, as shown in the work of \citet{yoon_speech_2020} and \citet{ahuja_style_2020}. 
These methods have significantly improved the naturalness of the generated gestures, but they are often limited to non-representational beats gestures, and offer limited flexibility in controlling the generated output \cite{ghorbani_zeroeggs_2022}.

In response to these limitations, recent research has increasingly focused on improving the adaptability of data-driven approaches. To address this, recent work has focused on increasing the adaptability of data-based approaches \cite{habibie_motion_2022, ghorbani_exemplar-based_2022, ghorbani_zeroeggs_2022, yoon_sgtoolkit_2021}. 
Techniques such as learning style embeddings for individual speakers \cite{ahuja_style_2020} and using variational frameworks to extract and blend motion styles from existing clips \cite{ghorbani_exemplar-based_2022} have been developed to provide greater control over gesture style and variation. 

Beyond the technical advancements in gesture synthesis, much work has also explored the impact of synthetic gestures on human-agent interaction. Studies have shown that users are more likely to engage with agents that use human-like gestures, which enhances the perceived personality of the agent \cite{neff_evaluating_2010, liu_two_2016} and facilitates relationship building \cite{gratch_creating_2007, bailenson_digital_2005}. Conversely, mismatched gestures can have a negative effect on interactions \cite{salem_generation_2012, wagner_gesture_2014, kelly_neural_2004}. While the primary effects of gestures are related to social perception and communication, there is limited evidence on their impact on user comprehension and task performance \cite{breckinridge_church_role_2007, davis_impact_2018}. Some studies suggest that gestures may aid learning, but more data is needed to draw definitive conclusions \cite{macedonia_imitation_2014, davis_impact_2018}.

Overall, the evolution of gesture generation methods from rule-based systems to sophisticated data-driven approaches has brought significant improvements in naturalness and adaptability. However, ongoing research continues to address the challenges of achieving flexible, contextually appropriate gesture synthesis that can effectively enhance human-agent interaction across various domains.

\section{Framework}

\subsection{AQGT}
The AQ-GT framework of \citet{voss2023aq} combines the strengths of Generative Adversarial Networks (GANs) \cite{goodfellow_generative_2014,creswell_generative_2018} and Vector Quantized Variational Autoencoders (VQ-VAE) \cite{oord_neural_2018}, specifically the improved VQ-VAE-2 model by \citet{razavi_generating_2019}, to process sequential data. This is achieved through a two-part approach, where four input modalities prior gestures, text, audio, and speaker identity are trained using a Gated Recurrent Unit (GRU) in combination with a transformer architecture \cite{cho_learning_2014,vaswani_attention_2023}.

The processing of each modality is distinct. For prior gestures, the VQ-VAE-2 was pretrained with a Wasserstein Generative Adversarial Network with Divergence penalty (WGAN-div) \cite{wu_wasserstein_2018} to learn a discrete latent space of concise gesture sequences. This model was then combined with a GAN to produce more diverse codebooks and stabilize the training process, as demonstrated by \citet{esser_taming_2021}. The resulting VQ-VAE-2\_wdiv model is used to generate prior gestures.

The audio input undergoes two additional processing methods: a Wav2Vec2 framework \cite{baevski_wav2vec_2020} extracts prosodic features from the raw audio data, and an onset encoder model by \citet{liang_seeg_2022} detects the onset of the audio and decouples the speech input into semantically relevant and irrelevant cues. The outputs of these methods are then used as input for a GRU transformer.

The text input is processed using the pre-trained BERT model \cite{devlin_bert_2019}, which converts it into tokens and generates a high-level representation. Speaker identity is predicted using a Multilayer Perceptron (MLP) network, with an embedding space added to the dataset for possible speaker identities.

To process unknown gesture styles, a MLP combined with the "Reparameterization Trick" by \citet{kingma_auto-encoding_2022} is used. The GRU transformer combines the abilities of the GRU to learn non-linear, localized dependencies and the ability of the transformer to capture broader global dependencies within the input sequence \cite{voss2023aq}. 

This output is then processed by a temporal aligner, which uses a sliding window approach, concatenating three frames with the current frame and the result of the previous iteration. A MLP generates a reconstructed vector, which is then used as input for the VQG model to generate a new gesture vector. This process repeats until three gesture vectors are combined into the final gesture vector, defining the gesture visible in the output video.

\subsection{AQGT-a}

The AQ-GT-A framework \cite{vos_augmented_2023} builds on the original AQ-GT framework with an additional input modality that captures information about form and meaning. This additional input is intended to enrich the semantic meaning of the generated gestures. To incorporate this new feature, an additional training set was created using the SAGA corpus \cite{lucking_bielefeld_2010, lucking_data-based_2013}, which consists of annotated human gestures in a video dataset.

The annotations included information on one-handed and two-handed gestures, as well as whether the gesture conveyed a relative position or an entity description. In addition, entities for nouns and adjectives were classified to provide more detailed information about the semantic content of the gestures. These annotations resulted in 17 categorical labels, which were used as form and meaning input.

To process this new input, an embedding layer was trained to generate a specific embedding for each label. The resulting embeddings were then processed by a Multilayer Perceptron (MLP) \cite{kingma_auto-encoding_2022}, resulting in a multivariate normal distribution in the latent space.

In addition, an Augmented Prediction Network (aPN) was developed to combine information about previous gestures with meaning and feature information. The aPN provides a probability distribution over the 16 possible label categories, allowing the generation of coherent gestures that are consistent with previous gestures. This network consisted of a Gated Recurrent Unit (GRU) trained on four input frames, using an embedding layer to predict labels for each frame.

The output of the MLP and the aPN were then combined into the fifth input channel of the GRU transformer. The overall architecture of AQ-GT-A remained similar to the original framework, with only the addition of a fifth input modality to the GRU transformer. The final output of AQ-GT-A retained the same shape as the AQ-GT framework.

\section{Study design}
\label{study_design}
The objective of this study was to investigate the capacity of frameworks to generate semantically meaningful gestures. To achieve this, we designed an evaluation study that considered the training data on which the frameworks were developed. Sentences similar to those used for training were included to assess performance, as the model's behavior is shaped by its training data \cite{hammoudeh2024training}. Additionally, sentences with varying degrees of similarity to the training data were incorporated to evaluate generalization capabilities.

The first group, Saga AQ-GT, Saga AQ-GT-a and Saga annotated, consists of sentences derived directly from the SAGA corpus, which focuses on directions given in a virtual city environment. The sentences in the "Saga AQ-GT" and "Saga AQ-GT-a" groups contain the original SAGA test sentences without annotations. In contrast, the "Saga annotated" group contains the same sentences with annotations for specific entities, such as doors and windows

The second group, close-to-Saga AQ-GT and close-to-Saga AQ-GT-a, contains newly created sentences similar to those found in the SAGA corpus, but without the use of SAGA-specific entities. Instead of entities such as doors and windows, proxies such as boxes and tables are used. Despite these differences, direction indicators and common adjectives similar to those in the SAGA corpus are retained. This group tests the ability of the frameworks to generalise contextually similar but different sentences.

The third group, Movement AQ-GT and Movement AQ-GT-a, shifts the focus from specific entities to sentences describing general movements or actions. This group includes a wide range of actions, from common ones like eating or throwing to less common ones like peeling. The sentences may also combine these actions with common objects or more specific items. The purpose of evaluating this group is to assess the ability of the frameworks to handle more general contexts involving actions that are primarily performed by humans.

\subsection{Concepts in the Evaluation Study}
\label{concepts}

The study involved six concepts, each aimed at testing the frameworks' abilities to generate gestures with specific semantic meanings. These concepts were defined as follows:

\begin{itemize}
    \item \textbf{Object}: A physical entity that can be perceived through sight and touch \cite{noauthor_object_nodate}. Sentences featuring a tangible object.
    \item \textbf{Direction}: The action or function of indicating a location or orientation \cite{noauthor_direction_nodate}. Sentences including direction indicators.
    \item \textbf{Negation}: A statement involving negative words (e.g., "no", "not", "never")  \cite{noauthor_negation_nodate}. Sentences featuring negative words.
    \item \textbf{Shape}: External form or contour \cite{noauthor_shape_nodate}. Sentences including a word describing the form of an object.
    \item \textbf{Size}: The magnitude or dimensions of an object or entity \cite{noauthor_size_nodate}. Sentences containing words indicating the size of an object.
    \item \textbf{Movement}: A change in physical location or an action executed by a subject \cite{noauthor_movement_nodate}. Sentences describing movement or actions.
\end{itemize}

The concepts Object, Size, Shape, and Movement focused on generating iconic gestures, the Negation concept targeted metaphoric gestures, and the Direction concept aimed at producing deictic gestures.

\subsection{Study Setup}
The study was conducted entirely online using the SoSciSurvey platform\footnote{\url{https://www.soscisurvey.de/}}, with participant recruitment and management facilitated through Prolific\footnote{\url{https://www.prolific.com/}}. Participants were randomly assigned to one of the seven experimental groups using the randomization tool provided by SoSciSurvey, which was configured to ensure an even distribution across the groups. Participants were compensated at a rate of £9 per hour, with payment contingent upon the successful completion of all but two attention checks.

\subsection{Study Procedure}

The study was divided into two parts, with participants receiving detailed instructions before each part. Initially, participants provided demographic information, including age, gender, and their Prolific ID.

In the first part of the study, participants were shown 10 silent video clips, with the sequence randomized for each participant to minimize sequence effects. Each video was presented on a separate page, accompanied by 4-point Likert scales designed to measure the recognition of specific concepts outlined in \ref{concepts}. The scale options ranged from "I was unable to recognize the concept at any point" to "I recognized the concept." Attention checks were placed after the third and seventh videos, requiring participants to answer each question before proceeding.

After completing the first part, participants were given new instructions for the second part, where they evaluated the human-likeness, helpfulness, synchronicity, and speech-reflectiveness of the gestures. This time, the same 10 videos were presented with audio, in a newly randomized order. Participants adjusted their audio settings using an example video provided at the beginning of this section.

For this part, a modified version of the HEMVIP questionnaire \cite{jonell_hemvip_2021} was used. The questionnaire employed a 4-point Likert scale, ranging from "Bad" to "Excellent," to encourage decisive responses and avoid neutral ratings. Attention checks were again included after the third and seventh videos, and participants had to complete all questions before moving on.

After the second part, participants were asked whether they experienced any distractions during the study and if they had followed the instructions as required. The study concluded with a brief summary and the provision of a Prolific code for compensation. Participants were then informed that they had completed the study and could close the window.

\subsection{Participants}

A total of 141 participants completed the study. Four participants were excluded from the analysis due to failing two attention checks, leaving 137 participants for statistical analysis. This final sample included 81 women, 55 men, and one individual who did not specify their gender. The participants ranged in age from 19 to 73 years, with an average age of 31 years.
The final distribution of participants was as follows: Saga AQ-GT (18 participants), Saga AQ-GT-a (20 participants), Saga annotated (21 participants), close-to-Saga AQ-GT (20 participants), close-to-Saga AQ-GT-a (19 participants), movement AQ-GT (19 participants), and movement AQ-GT-a (20 participants). 

\begin{figure}
    \centering
    \includegraphics[width=0.9\linewidth]{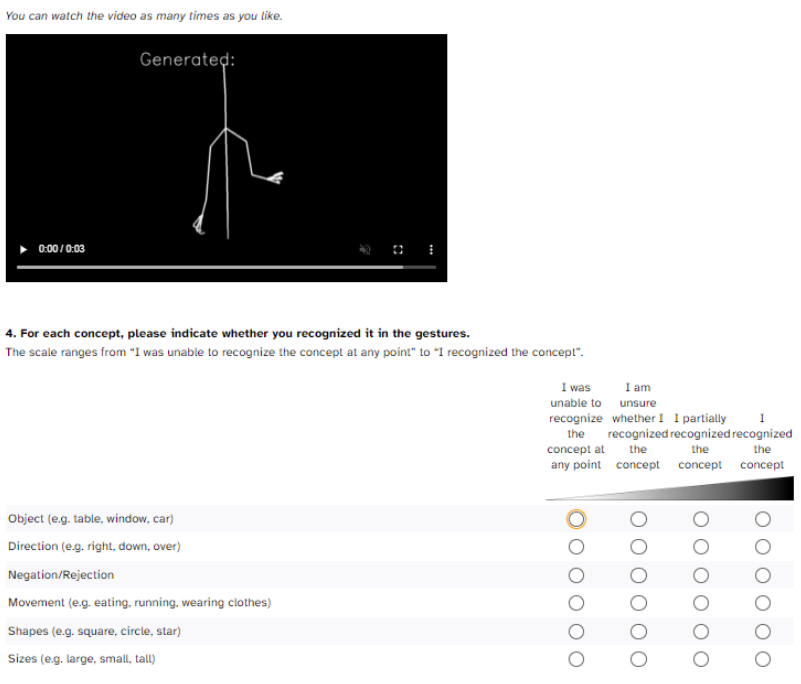}
    \caption{Example view of stage one of the evaluation study}\label{fig:setup}
\end{figure}

\begin{figure}
    \centering
    \includegraphics[width=0.9\linewidth]{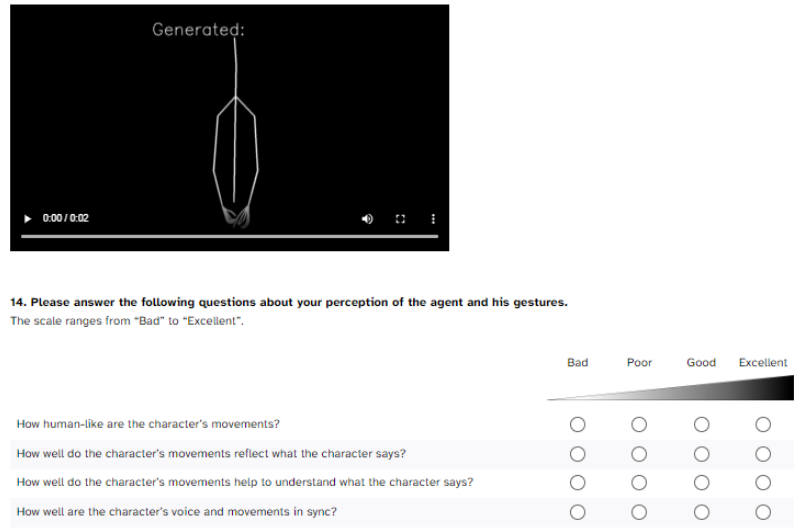}
    \caption{Example view of stage two of the evaluation study}\label{fig:setup_2}
\end{figure}

\section{Results}

\begin{figure}
    \centering
    \includegraphics[width=0.9\linewidth]{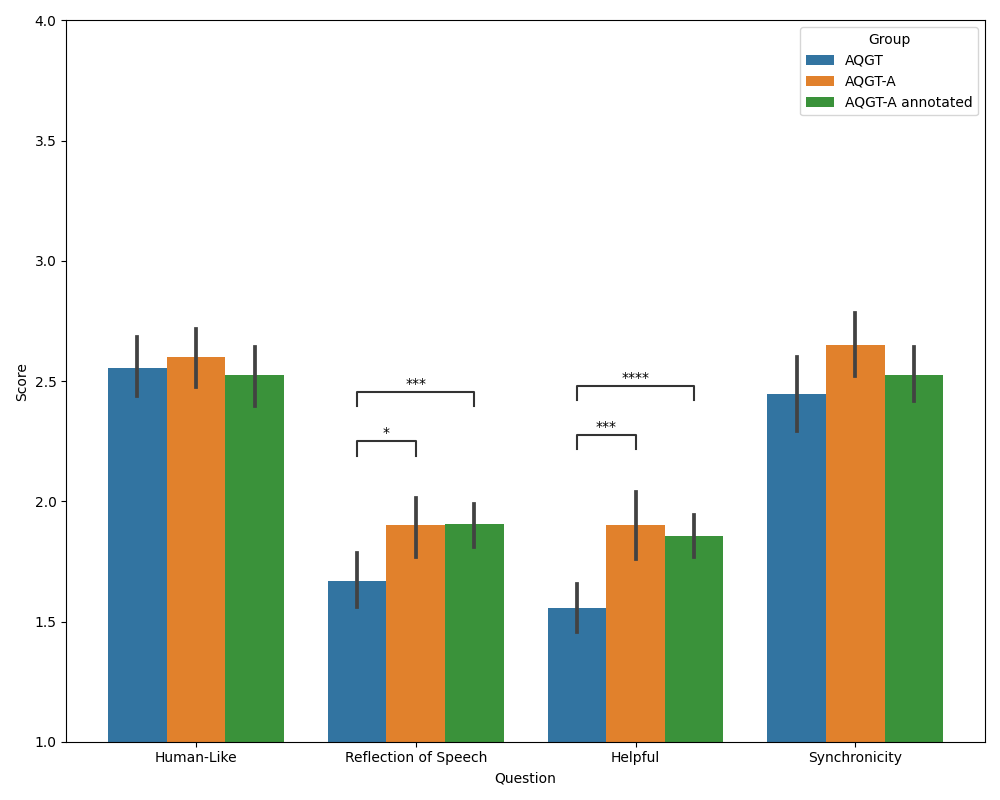}
    \caption{Comparing scores across four evaluation dimensions: Human-Like, Reflection of Speech, Helpful, and Synchronicity for three experimental groups: AQGT, AQGT-A, and AQGT-A Annotated. The y-axis represents the score, ranging from 1.0 to 4.0, with error bars indicating standard deviations. }\label{fig:results_hemvip}
\end{figure}

To determine whether the data met the assumptions for parametric testing, we performed a Shapiro-Wilk test for normality. The results showed significant deviations from normality in all conditions, indicating that the data were not normally distributed. Given this violation of normality, non-parametric statistical methods were used.
To compare differences between conditions, Mann-Whitney U tests were performed. As multiple comparisons were performed, we also applied a Bonferroni correction to adjust for inflated type I error rates, thus ensuring a more conservative threshold for statistical significance.

An analysis of the HEMVIP questionnaire results, presented in Figure \ref{fig:results_hemvip}, indicated varied perceptions across the groups. For Helpfulness, ratings for AQGT-A (M = 1.90, SD = 0.95) and AQGT-A annotated (M = 1.86, SD = 0.64) were higher than for AQGT (M = 1.56, SD = 0.69). The differences were significant between AQGT and AQGT-A (p = 0.001) and between AQGT and AQGT-A annotated (p = 0.00001). In contrast, ratings for Human-Likeness were similar across all three conditions: AQGT (M = 2.56, SD = 0.83), AQGT-A (M = 2.60, SD = 0.86), and AQGT-A annotated (M = 2.52, SD = 0.91), with no significant differences reported. For Reflection of Speech, both AQGT-A (M = 1.90, SD = 0.89) and AQGT-A annotated (M = 1.90, SD = 0.69) scored higher than AQGT (M = 1.67, SD = 0.75). These differences were statistically significant (p = 0.0153 and p = 0.0004, respectively). Lastly, Synchronicity scores were comparable across AQGT (M = 2.44, SD = 1.07), AQGT-A (M = 2.65, SD = 0.91), and AQGT-A annotated (M = 2.52, SD = 0.85), with no significant differences found.

\begin{figure}
    \centering
    \includegraphics[width=0.9\linewidth]{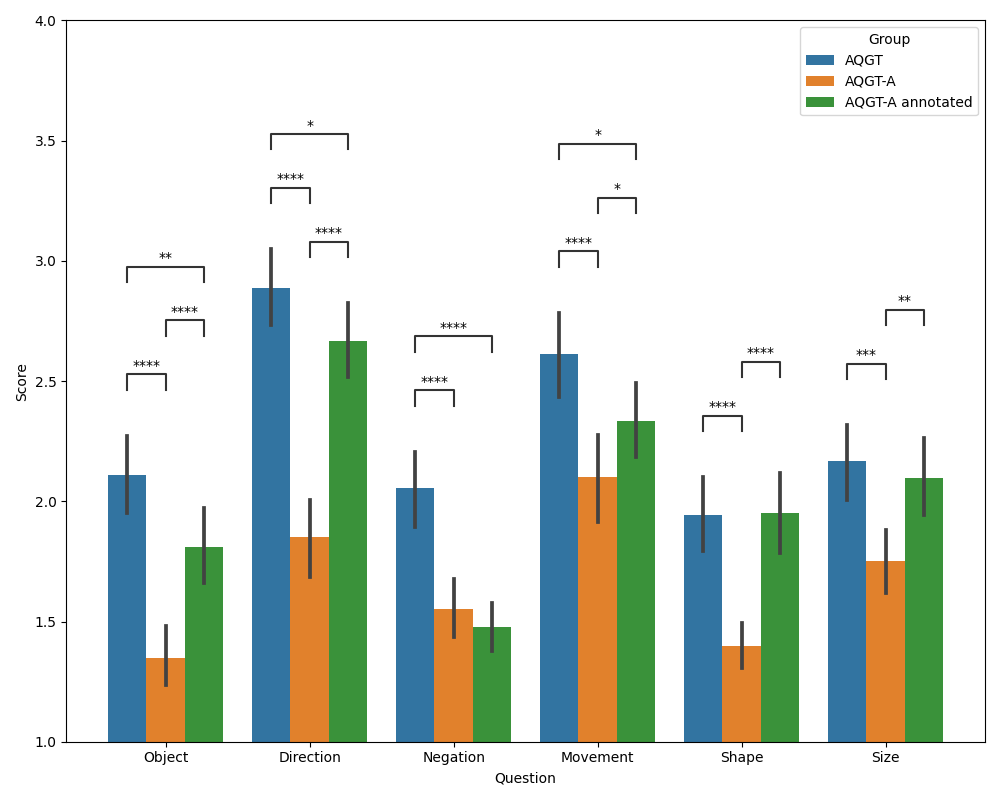}
    \caption{Comparison of the scores from the six concepts on sentences derived directly from the SAGA corpus. See \ref{study_design} for more information. The asterisks denote the statistical significance levels (p < 0.05, **p < 0.001, ***p < 0.0001)}
    \label{fig:results_saga}
\end{figure}

Regarding the results for sentences from the SAGA corpus, shown in Figure \ref{fig:results_saga}, significant differences were found across most categories. For Object recognition, AQGT (M = 2.11, SD = 1.10) outperformed both AQGT-A (M = 1.35, SD = 0.91) and AQGT-A annotated (M = 1.81, SD = 1.10). All pairwise comparisons were significant (AQGT vs AQGT-A, p = 0.0000; AQGT vs AQGT-A annotated, p = 0.0062; AQGT-A vs AQGT-A annotated, p = 0.0000). A similar pattern was observed for Direction, where AQGT (M = 2.89, SD = 1.10) was rated highest compared to AQGT-A (M = 1.85, SD = 1.16) and AQGT-A annotated (M = 2.67, SD = 1.13), with all comparisons yielding significant differences (AQGT vs AQGT-A, p = 0.0000; AQGT vs AQGT-A annotated, p = 0.0436; AQGT-A vs AQGT-A annotated, p = 0.0000). For Negation, AQGT (M = 2.06, SD = 1.13) scored significantly higher than both AQGT-A (M = 1.55, SD = 0.87; p = 0.0000) and AQGT-A annotated (M = 1.48, SD = 0.73; p = 0.0000). In the Movement category, AQGT (M = 2.61, SD = 1.16) again received the highest scores, with all pairwise comparisons to AQGT-A (M = 2.10, SD = 1.30) and AQGT-A annotated (M = 2.33, SD = 1.13) proving significant (p = 0.0000, p = 0.0155, and p = 0.0244, respectively). For Shape, AQGT (M = 1.94, SD = 1.08) and AQGT-A annotated (M = 1.95, SD = 1.22) were rated significantly higher than AQGT-A (M = 1.40, SD = 0.66; p = 0.0000 for both comparisons). Similarly, for Size, AQGT (M = 2.17, SD = 1.12) was rated significantly higher than AQGT-A (M = 1.75, SD = 1.00; p = 0.0001), and AQGT-A annotated (M = 2.10, SD = 1.19) was rated significantly higher than AQGT-A (p = 0.0026).

\begin{figure}
    \centering
    \includegraphics[width=0.9\linewidth]{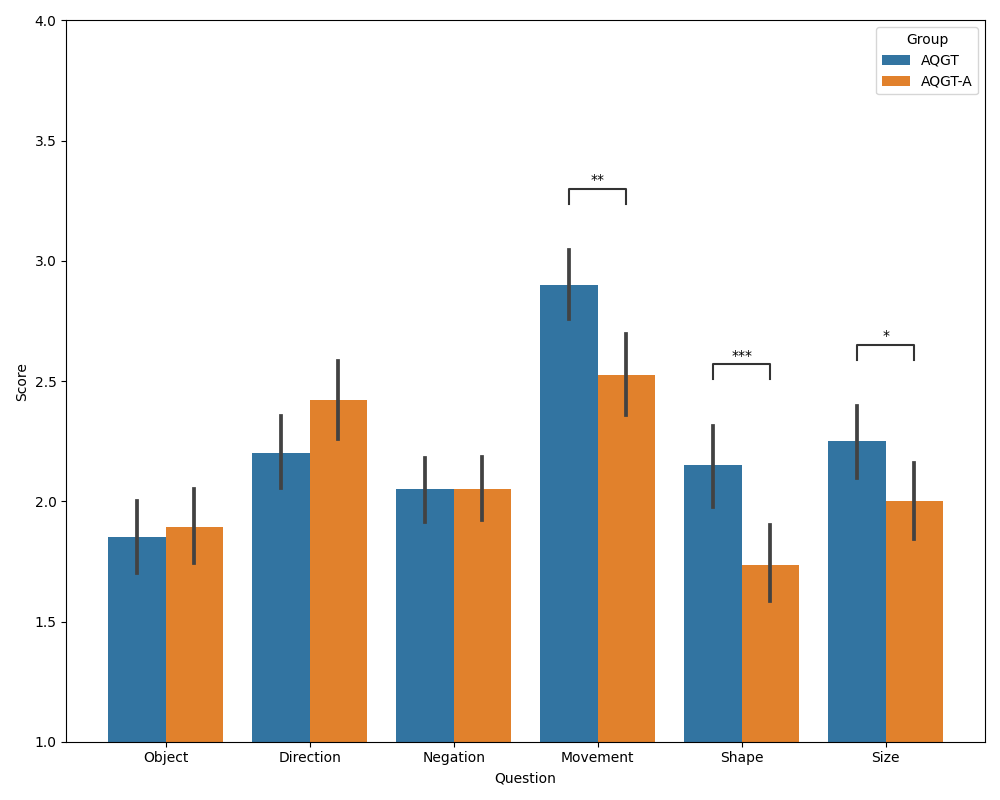}
    \caption{Comparison of the scores from the six concepts on sentences similar to those in the saga corpus. See \ref{study_design} for more information. The asterisks denote the statistical significance levels (p < 0.05, **p < 0.001, ***p < 0.0001)}
    \label{fig:results_close}
\end{figure}

An analysis of sentences close to the SAGA corpus, shown in Figure \ref{fig:results_close}, revealed fewer significant differences. For Movement, AQGT (M = 2.90, SD = 1.05) was rated significantly higher than AQGT-A (M = 2.53, SD = 1.23; p = 0.0044). For Shape, AQGT (M = 2.15, SD = 1.20) also scored significantly higher than AQGT-A (M = 1.74, SD = 1.12; p = 0.0003). A significant difference was also found for Size, where AQGT (M = 2.25, SD = 1.09) again outperformed AQGT-A (M = 2.00, SD = 1.13; p = 0.0183). No significant differences were found for the Object, Direction, or Negation categories.

\begin{figure}
    \centering
    \includegraphics[width=0.9\linewidth]{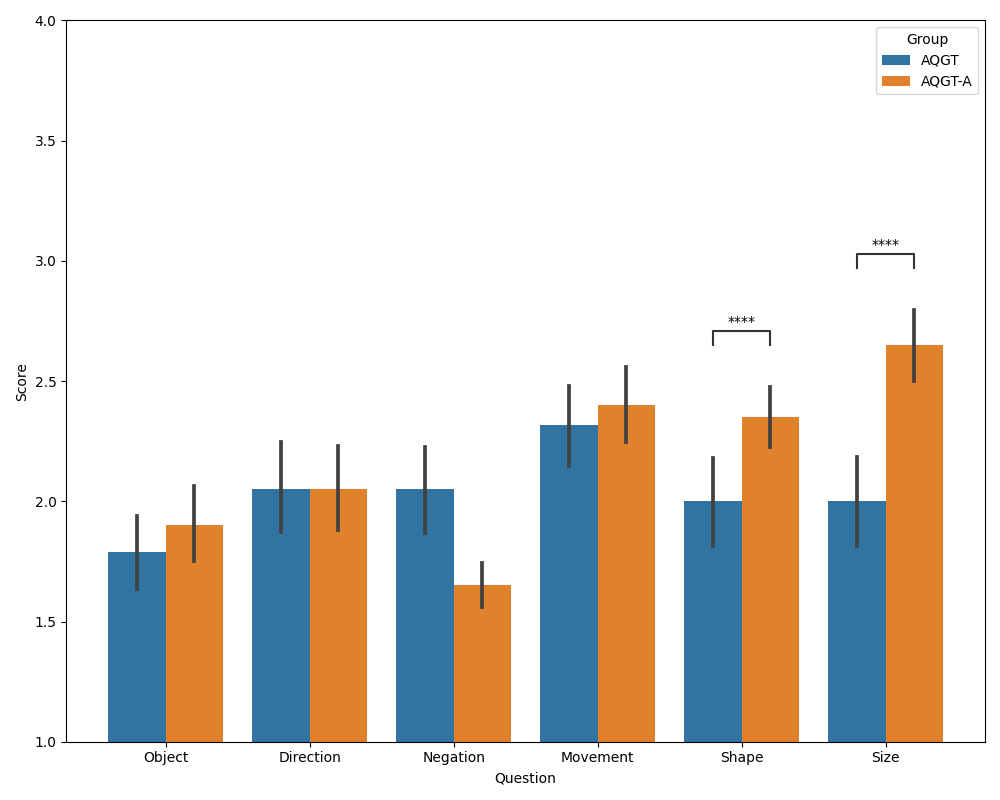}
    \caption{Comparison of the Likert scale score from the six concepts on sentences that describe general movements or actions. See \ref{study_design} for more information The asterisks denote the statistical significance levels (p < 0.05, **p < 0.001, ***p < 0.0001)}
    \label{fig:results_movement}
\end{figure}

Finally, for the movement-focused sentences detailed in Figure \ref{fig:results_movement}, a different pattern of results emerged. For Shape, AQGT-A (M = 2.35, SD = 0.91) was rated significantly higher than AQGT (M = 2.00, SD = 1.22; p = 0.0000). A similar result was found for Size, where AQGT-A (M = 2.65, SD = 1.06) also scored significantly higher than AQGT (M = 2.00, SD = 1.30; p = 0.0000). No significant differences were reported for the Object, Direction, Negation, or Movement categories.

\section{Discussion}

This study aimed to evaluate the ability of two synthetic gesture generation frameworks (AQ-GT and AQ-GT-A) to convey semantic meaning and assess human perception of the generated gestures. The results paint a complex picture, suggesting that including explicit semantic annotations does not necessarily improve concept recognition, and that the context of the utterance is crucial for the effectiveness of the generated gestures.

Our first hypothesis (H1) predicted that the gestures produced by AQ-GT-A would be more semantically rich than those produced by AQ-GT. However, the results largely contradicted this. For sentences taken directly from the SAGA corpus, the original AQ-GT framework, which did not include any explicit semantic input, was significantly more effective at conveying concepts across almost all categories, including Object, Direction, Negation, Movement, Shape and Size. This finding was unexpected, given that AQ-GT-a was specifically designed to improve semantic representation by incorporating additional form and meaning features from the SAGA corpus. One possible explanation is that the explicit annotations in AQ-GT-a constrained the model's output, resulting in gestures that were more ambiguous or less expressive than those implicitly learned by AQ-GT from the raw data. It is possible that the original model developed a more robust, albeit implicit, mapping between prosodic and textual cues and gestural representation, which participants found more intuitive in this specific context.

The sub-hypotheses that (H1.2 and H1.3) AQ-GT-a with and without annotations would be more semantically rich, and that (H1.4, H1.5 and H1.6) the inclusion of SAGA-Entities would enhance recognisability, were not supported in the SAGA and close-to-SAGA contexts. In fact, for SAGA sentences, the annotated AQ-GT-a condition, which contained the most explicit semantic information, often performed better than unannotated AQ-GT-a, yet was still consistently outperformed by original AQ-GT. This suggests that the method of incorporating semantic labels, as implemented in AQ-GT-a, may not be optimal for sentences that closely resemble the training data. However, for movement-focused sentences representing a shift away from the training domain, AQ-GT-a performed significantly better than AQ-GT in representing shape and size. This finding suggests that the model's generalisation capabilities may be enhanced by semantic annotations when faced with novel contexts.

Our second hypothesis (H2) examined how people perceive the generated gestures. In this case, the results were more aligned with our predictions. We predicted that AQ-GT-A gestures would be perceived as more human-like, synchronous and expressive. Although there were no significant differences in perceived human-likeness or synchronicity across any conditions, significant differences in expressiveness were observed, as measured by helpfulness and reflection of speech. Both AQ-GT-a and AQ-GT-a with annotations were rated as significantly more helpful and reflective of speech than AQ-GT. This supports sub-hypotheses H2.1 and H2.2, indicating that semantic augmentation in AQ-GT-a, with or without annotations, successfully enhances the perceived communicative quality of gestures without making them appear more human-like. Distinguishing between naturalness and communicative effectiveness is important because it suggests that users can differentiate between gestures that appear physically realistic and those that effectively support the spoken message.

Further analysis of the second hypothesis revealed that the prediction that the SAGA context with SAGA entities (annotated) would be more expressive (H2.5) was supported by significantly higher ratings for 'Helpful' and 'Reflection of Speech' compared to the base AQ-GT model. However, the expectation that gestures would be less expressive when SAGA entities were excluded from the context (H2.6) was not entirely confirmed, as unannotated AQ-GT-a still outperformed original AQ-GT in terms of expressiveness. This suggests that the underlying architectural changes in AQ-GT-a contribute to expressiveness even in the absence of explicit entity annotation for a given sentence. Finally, as predicted in H2.7, the human-likeness and synchronicity ratings were similar across different contexts, which reinforces the idea that semantic augmentations primarily impact the communicative function rather than the perceived naturalness of motion. This highlights a critical trade-off: specialising a model with explicit semantics (as in AQ-GT-a) may enhance its ability to generalise to new semantic domains (e.g. movement, shape and size) and improve perceived expressiveness, but this may come at the cost of clarity of concepts within its original training domain. The simpler AQ-GT model's implicit learning proved more effective for the specific SAGA context it was trained on.

\section{Limitations}
While this study provides valuable insights into the performance of the AQ-GT and AQ-GT-a gesture generation frameworks, several limitations should be considered when interpreting the results.
First, the evaluation was conducted using a specific set of sentences and concepts. The three contexts (SAGA, close-to-SAGA, and Movement) were designed to test performance within and outside the original training domain. However, the scope of language and gestures remains relatively narrow. The SAGA corpus is focused on giving directions in a virtual city, and the novel sentences, while different, may not fully represent the broad spectrum of human communication.
Second, the study participants were recruited through an online platform. While this allows for a diverse sample in terms of age and gender, it introduces potential variability in the viewing environment, such as screen size, audio quality, and potential distractions, which could have influenced participants' perceptions. 
Finally, the evaluation of concept recognition relied on self-reported ratings from participants using Likert scales. This method provides a measure of perceived clarity, but it is subjective and may not perfectly correlate with the objective presence or accuracy of a gestural representation.

\section{Conclusion} 

In conclusion, this study provides a nuanced view on the role of explicit semantic information in data-driven gesture generation. The surprising outperformance of the AQ-GT model on its native SAGA corpus sentences suggests that adding semantic annotations is not a simple solution and can sometimes hinder performance, possibly by overly constraining the model. However, the improved performance of AQ-GT-a on out-of-domain sentences highlights the potential for such annotations to improve model generalization. Furthermore, the separation of perceived expressiveness from human-likeness indicates that the communicative function of gestures can be enhanced independently of their physical realism. Future research should focus on refining how semantic information is integrated into generation models and on developing more comprehensive evaluation methodologies that capture the multifaceted impact of gestures on human-agent interaction.

\clearpage
\bibliographystyle{ACM-Reference-Format}
\bibliography{references}


\begin{thebibliography}{50}


\ifx \showCODEN    \undefined \def \showCODEN     #1{\unskip}     \fi
\ifx \showDOI      \undefined \def \showDOI       #1{#1}\fi
\ifx \showISBNx    \undefined \def \showISBNx     #1{\unskip}     \fi
\ifx \showISBNxiii \undefined \def \showISBNxiii  #1{\unskip}     \fi
\ifx \showISSN     \undefined \def \showISSN      #1{\unskip}     \fi
\ifx \showLCCN     \undefined \def \showLCCN      #1{\unskip}     \fi
\ifx \shownote     \undefined \def \shownote      #1{#1}          \fi
\ifx \showarticletitle \undefined \def \showarticletitle #1{#1}   \fi
\ifx \showURL      \undefined \def \showURL       {\relax}        \fi
\providecommand\bibfield[2]{#2}
\providecommand\bibinfo[2]{#2}
\providecommand\natexlab[1]{#1}
\providecommand\showeprint[2][]{arXiv:#2}

\bibitem[noa(2004)]%
        {noauthor_beat_nodate}
 \bibinfo{year}{2004}\natexlab{}.
\newblock \bibinfo{booktitle}{\emph{{BEAT}: the Behavior Expression Animation Toolkit {\textbar} {SpringerLink}}}.
\newblock
\urldef\tempurl%
\url{https://link.springer.com/chapter/10.1007/978-3-662-08373-4_8}
\showURL{%
\tempurl}


\bibitem[Ahuja et~al\mbox{.}(2020)]%
        {ahuja_style_2020}
\bibfield{author}{\bibinfo{person}{Chaitanya Ahuja}, \bibinfo{person}{Dong~Won Lee}, \bibinfo{person}{Yukiko~I. Nakano}, {and} \bibinfo{person}{Louis-Philippe Morency}.} \bibinfo{year}{2020}\natexlab{}.
\newblock \bibinfo{title}{Style Transfer for Co-Speech Gesture Animation: A Multi-Speaker Conditional-Mixture Approach}.
\newblock
\newblock
\urldef\tempurl%
\url{https://doi.org/10.48550/arXiv.2007.12553}
\showDOI{\tempurl}
\showeprint[arxiv]{2007.12553 [cs]}


\bibitem[Alexanderson et~al\mbox{.}({[n.\,d.]})]%
        {alexanderson_style-controllable_2020}
\bibfield{author}{\bibinfo{person}{Simon Alexanderson}, \bibinfo{person}{Gustav~Eje Henter}, \bibinfo{person}{Taras Kucherenko}, {and} \bibinfo{person}{Jonas Beskow}.} \bibinfo{year}{[n.\,d.]}\natexlab{}.
\newblock \showarticletitle{Style-Controllable Speech-Driven Gesture Synthesis Using Normalising Flows}. In \bibinfo{booktitle}{\emph{Computer Graphics Forum}} (2020), Vol.~\bibinfo{volume}{39}. \bibinfo{publisher}{Wiley Online Library}, \bibinfo{pages}{487--496}.
\newblock
\newblock
\shownote{Issue: 2}.


\bibitem[Baevski et~al\mbox{.}({[n.\,d.]})]%
        {baevski_wav2vec_2020}
\bibfield{author}{\bibinfo{person}{Alexei Baevski}, \bibinfo{person}{Henry Zhou}, \bibinfo{person}{Abdelrahman Mohamed}, {and} \bibinfo{person}{Michael Auli}.} \bibinfo{year}{[n.\,d.]}\natexlab{}.
\newblock \bibinfo{title}{wav2vec 2.0: A Framework for Self-Supervised Learning of Speech Representations}.
\newblock
\newblock
\urldef\tempurl%
\url{https://doi.org/10.48550/arXiv.2006.11477}
\showDOI{\tempurl}
\showeprint[arxiv]{2006.11477 [cs, eess]}


\bibitem[Bailenson and Yee({[n.\,d.]})]%
        {bailenson_digital_2005}
\bibfield{author}{\bibinfo{person}{Jeremy~N. Bailenson} {and} \bibinfo{person}{Nick Yee}.} \bibinfo{year}{[n.\,d.]}\natexlab{}.
\newblock \showarticletitle{Digital Chameleons: Automatic Assimilation of Nonverbal Gestures in Immersive Virtual Environments}.
\newblock  \bibinfo{volume}{16}, \bibinfo{number}{10} (\bibinfo{year}{[n.\,d.]}), \bibinfo{pages}{814--819}.
\newblock
\showISSN{1467-9280}
\urldef\tempurl%
\url{https://doi.org/10.1111/j.1467-9280.2005.01619.x}
\showDOI{\tempurl}
\newblock
\shownote{Place: United Kingdom Publisher: Blackwell Publishing}.


\bibitem[Bergmann and Kopp({[n.\,d.]})]%
        {ruttkay_gnetic_2009}
\bibfield{author}{\bibinfo{person}{Kirsten Bergmann} {and} \bibinfo{person}{Stefan Kopp}.} \bibinfo{year}{[n.\,d.]}\natexlab{}.
\newblock \showarticletitle{{GNetIc} – Using Bayesian Decision Networks for Iconic Gesture Generation}.
\newblock In \bibinfo{booktitle}{\emph{Intelligent Virtual Agents}}, \bibfield{editor}{\bibinfo{person}{Zsófia Ruttkay}, \bibinfo{person}{Michael Kipp}, \bibinfo{person}{Anton Nijholt}, {and} \bibinfo{person}{Hannes~Högni Vilhjálmsson}} (Eds.). Vol.~\bibinfo{volume}{5773}. \bibinfo{publisher}{Springer Berlin Heidelberg}, \bibinfo{pages}{76--89}.
\newblock
\showISBNx{978-3-642-04379-6 978-3-642-04380-2}
\urldef\tempurl%
\url{https://doi.org/10.1007/978-3-642-04380-2_12}
\showDOI{\tempurl}
\newblock
\shownote{Series Title: Lecture Notes in Computer Science}.


\bibitem[Breckinridge~Church et~al\mbox{.}({[n.\,d.]})]%
        {breckinridge_church_role_2007}
\bibfield{author}{\bibinfo{person}{Ruth Breckinridge~Church}, \bibinfo{person}{Philip Garber}, {and} \bibinfo{person}{Kathryn Rogalski}.} \bibinfo{year}{[n.\,d.]}\natexlab{}.
\newblock \showarticletitle{The role of gesture in memory and social communication}.
\newblock  \bibinfo{volume}{7}, \bibinfo{number}{2} (\bibinfo{year}{[n.\,d.]}), \bibinfo{pages}{137--158}.
\newblock
\showISSN{1568-1475, 1569-9773}
\urldef\tempurl%
\url{https://doi.org/10.1075/gest.7.2.02bre}
\showDOI{\tempurl}


\bibitem[Cho et~al\mbox{.}({[n.\,d.]})]%
        {cho_learning_2014}
\bibfield{author}{\bibinfo{person}{Kyunghyun Cho}, \bibinfo{person}{Bart van Merrienboer}, \bibinfo{person}{Caglar Gulcehre}, \bibinfo{person}{Dzmitry Bahdanau}, \bibinfo{person}{Fethi Bougares}, \bibinfo{person}{Holger Schwenk}, {and} \bibinfo{person}{Yoshua Bengio}.} \bibinfo{year}{[n.\,d.]}\natexlab{}.
\newblock \bibinfo{title}{Learning Phrase Representations using {RNN} Encoder-Decoder for Statistical Machine Translation}.
\newblock
\newblock
\urldef\tempurl%
\url{https://doi.org/10.48550/arXiv.1406.1078}
\showDOI{\tempurl}
\showeprint[arxiv]{1406.1078 [cs, stat]}


\bibitem[Creswell et~al\mbox{.}({[n.\,d.]})]%
        {creswell_generative_2018}
\bibfield{author}{\bibinfo{person}{Antonia Creswell}, \bibinfo{person}{Tom White}, \bibinfo{person}{Vincent Dumoulin}, \bibinfo{person}{Kai Arulkumaran}, \bibinfo{person}{Biswa Sengupta}, {and} \bibinfo{person}{Anil~A. Bharath}.} \bibinfo{year}{[n.\,d.]}\natexlab{}.
\newblock \showarticletitle{Generative Adversarial Networks: An Overview}.
\newblock  \bibinfo{volume}{35}, \bibinfo{number}{1} (\bibinfo{year}{[n.\,d.]}), \bibinfo{pages}{53--65}.
\newblock
\showISSN{1053-5888, 1558-0792}
\urldef\tempurl%
\url{https://doi.org/10.1109/MSP.2017.2765202}
\showDOI{\tempurl}
\showeprint[arxiv]{1710.07035 [cs]}


\bibitem[Davis({[n.\,d.]})]%
        {davis_impact_2018}
\bibfield{author}{\bibinfo{person}{Robert~O. Davis}.} \bibinfo{year}{[n.\,d.]}\natexlab{}.
\newblock \showarticletitle{The impact of pedagogical agent gesturing in multimedia learning environments: A meta-analysis}.
\newblock   \bibinfo{volume}{24} (\bibinfo{year}{[n.\,d.]}), \bibinfo{pages}{193--209}.
\newblock
\showISSN{1747-938X}
\urldef\tempurl%
\url{https://doi.org/10.1016/j.edurev.2018.05.002}
\showDOI{\tempurl}


\bibitem[Devlin et~al\mbox{.}({[n.\,d.]})]%
        {devlin_bert_2019}
\bibfield{author}{\bibinfo{person}{Jacob Devlin}, \bibinfo{person}{Ming-Wei Chang}, \bibinfo{person}{Kenton Lee}, {and} \bibinfo{person}{Kristina Toutanova}.} \bibinfo{year}{[n.\,d.]}\natexlab{}.
\newblock \bibinfo{title}{{BERT}: Pre-training of Deep Bidirectional Transformers for Language Understanding}.
\newblock
\newblock
\urldef\tempurl%
\url{https://doi.org/10.48550/arXiv.1810.04805}
\showDOI{\tempurl}
\showeprint[arxiv]{1810.04805 [cs]}


\bibitem[Esser et~al\mbox{.}({[n.\,d.]})]%
        {esser_taming_2021}
\bibfield{author}{\bibinfo{person}{Patrick Esser}, \bibinfo{person}{Robin Rombach}, {and} \bibinfo{person}{Björn Ommer}.} \bibinfo{year}{[n.\,d.]}\natexlab{}.
\newblock \bibinfo{title}{Taming Transformers for High-Resolution Image Synthesis}.
\newblock
\newblock
\urldef\tempurl%
\url{https://doi.org/10.48550/arXiv.2012.09841}
\showDOI{\tempurl}
\showeprint[arxiv]{2012.09841 [cs]}


\bibitem[Ferstl and McDonnell(2018)]%
        {ferstlInvestigatingUseRecurrent2018a}
\bibfield{author}{\bibinfo{person}{Ylva Ferstl} {and} \bibinfo{person}{Rachel McDonnell}.} \bibinfo{year}{2018}\natexlab{}.
\newblock \showarticletitle{Investigating the use of recurrent motion modelling for speech gesture generation}. In \bibinfo{booktitle}{\emph{Proceedings of the 18th {International} {Conference} on {Intelligent} {Virtual} {Agents}}}. \bibinfo{publisher}{ACM}, \bibinfo{address}{Sydney NSW Australia}, \bibinfo{pages}{93--98}.
\newblock
\showISBNx{978-1-4503-6013-5}
\urldef\tempurl%
\url{https://doi.org/10.1145/3267851.3267898}
\showDOI{\tempurl}


\bibitem[Ghorbani et~al\mbox{.}({[n.\,d.]})]%
        {ghorbani_exemplar-based_2022}
\bibfield{author}{\bibinfo{person}{Saeed Ghorbani}, \bibinfo{person}{Ylva Ferstl}, {and} \bibinfo{person}{Marc-André Carbonneau}.} \bibinfo{year}{[n.\,d.]}\natexlab{}.
\newblock \showarticletitle{Exemplar-based Stylized Gesture Generation from Speech: An Entry to the {GENEA} Challenge 2022}. In \bibinfo{booktitle}{\emph{Proceedings of the 2022 International Conference on Multimodal Interaction}} (Bengaluru India, 2022-11-07). \bibinfo{publisher}{{ACM}}, \bibinfo{pages}{778--783}.
\newblock
\showISBNx{978-1-4503-9390-4}
\urldef\tempurl%
\url{https://doi.org/10.1145/3536221.3558068}
\showDOI{\tempurl}


\bibitem[Ghorbani et~al\mbox{.}(2022)]%
        {ghorbani_zeroeggs_2022}
\bibfield{author}{\bibinfo{person}{Saeed Ghorbani}, \bibinfo{person}{Ylva Ferstl}, \bibinfo{person}{Daniel Holden}, \bibinfo{person}{Nikolaus~F. Troje}, {and} \bibinfo{person}{Marc-André Carbonneau}.} \bibinfo{year}{2022}\natexlab{}.
\newblock \bibinfo{title}{{ZeroEGGS}: Zero-shot Example-based Gesture Generation from Speech}.
\newblock
\newblock
\urldef\tempurl%
\url{https://doi.org/10.48550/arXiv.2209.07556}
\showDOI{\tempurl}
\showeprint[arxiv]{2209.07556 [cs]}


\bibitem[Goodfellow et~al\mbox{.}({[n.\,d.]})]%
        {goodfellow_generative_2014}
\bibfield{author}{\bibinfo{person}{Ian~J. Goodfellow}, \bibinfo{person}{Jean Pouget-Abadie}, \bibinfo{person}{Mehdi Mirza}, \bibinfo{person}{Bing Xu}, \bibinfo{person}{David Warde-Farley}, \bibinfo{person}{Sherjil Ozair}, \bibinfo{person}{Aaron Courville}, {and} \bibinfo{person}{Yoshua Bengio}.} \bibinfo{year}{[n.\,d.]}\natexlab{}.
\newblock \bibinfo{title}{Generative Adversarial Networks}.
\newblock
\newblock
\urldef\tempurl%
\url{https://doi.org/10.48550/arXiv.1406.2661}
\showDOI{\tempurl}
\showeprint[arxiv]{1406.2661 [cs, stat]}


\bibitem[Gratch et~al\mbox{.}(2007)]%
        {gratch_creating_2007}
\bibfield{author}{\bibinfo{person}{Jonathan Gratch}, \bibinfo{person}{Ning Wang}, \bibinfo{person}{Jillian Gerten}, \bibinfo{person}{Edward Fast}, {and} \bibinfo{person}{Robin Duffy}.} \bibinfo{year}{2007}\natexlab{}.
\newblock \showarticletitle{Creating Rapport with Virtual Agents}.
\newblock In \bibinfo{booktitle}{\emph{Lecture Notes in Artificial Intelligence; Proceedings of the 7th International Conference on Intelligent Virtual Agents ({IVA})}}.
\newblock
\urldef\tempurl%
\url{http://ict.usc.edu/pubs/Creating%20Rapport%20with%20Virtual%20Agents.pdf}
\showURL{%
\tempurl}


\bibitem[Habibie et~al\mbox{.}({[n.\,d.]})]%
        {habibie_motion_2022}
\bibfield{author}{\bibinfo{person}{Ikhsanul Habibie}, \bibinfo{person}{Mohamed Elgharib}, \bibinfo{person}{Kripashindu Sarkar}, \bibinfo{person}{Ahsan Abdullah}, \bibinfo{person}{Simbarashe Nyatsanga}, {and} \bibinfo{person}{Christian Theobalt}.} \bibinfo{year}{[n.\,d.]}\natexlab{}.
\newblock \showarticletitle{A Motion Matching-based Framework for Controllable Gesture Synthesis from Speech}.
\newblock
\urldef\tempurl%
\url{https://vcai.mpi-inf.mpg.de/projects/SpeechGestureMatching/}
\showURL{%
\tempurl}


\bibitem[Hammoudeh and Lowd(2024)]%
        {hammoudeh2024training}
\bibfield{author}{\bibinfo{person}{Zayd Hammoudeh} {and} \bibinfo{person}{Daniel Lowd}.} \bibinfo{year}{2024}\natexlab{}.
\newblock \showarticletitle{Training data influence analysis and estimation: A survey}.
\newblock \bibinfo{journal}{\emph{Machine Learning}} \bibinfo{volume}{113}, \bibinfo{number}{5} (\bibinfo{year}{2024}), \bibinfo{pages}{2351--2403}.
\newblock


\bibitem[Henter et~al\mbox{.}(2020)]%
        {henterMoGlowProbabilisticControllable2020}
\bibfield{author}{\bibinfo{person}{Gustav~Eje Henter}, \bibinfo{person}{Simon Alexanderson}, {and} \bibinfo{person}{Jonas Beskow}.} \bibinfo{year}{2020}\natexlab{}.
\newblock \showarticletitle{{MoGlow}: {Probabilistic} and controllable motion synthesis using normalising flows}.
\newblock \bibinfo{journal}{\emph{ACM Transactions on Graphics}} \bibinfo{volume}{39}, \bibinfo{number}{6} (\bibinfo{date}{Dec.} \bibinfo{year}{2020}), \bibinfo{pages}{1--14}.
\newblock
\showISSN{0730-0301, 1557-7368}
\urldef\tempurl%
\url{https://doi.org/10.1145/3414685.3417836}
\showDOI{\tempurl}
\newblock
\shownote{arXiv:1905.06598 [cs, eess, stat]}.


\bibitem[Jonell et~al\mbox{.}(2021)]%
        {jonell_hemvip_2021}
\bibfield{author}{\bibinfo{person}{Patrik Jonell}, \bibinfo{person}{Youngwoo Yoon}, \bibinfo{person}{Pieter Wolfert}, \bibinfo{person}{Taras Kucherenko}, {and} \bibinfo{person}{Gustav~Eje Henter}.} \bibinfo{year}{2021}\natexlab{}.
\newblock \showarticletitle{{HEMVIP}: {Human} {Evaluation} of {Multiple} {Videos} in {Parallel}}. In \bibinfo{booktitle}{\emph{Proceedings of the 2021 {International} {Conference} on {Multimodal} {Interaction}}}. \bibinfo{publisher}{ACM}, \bibinfo{address}{Montréal QC Canada}, \bibinfo{pages}{707--711}.
\newblock
\showISBNx{978-1-4503-8481-0}
\urldef\tempurl%
\url{https://doi.org/10.1145/3462244.3479957}
\showDOI{\tempurl}


\bibitem[Kelly et~al\mbox{.}({[n.\,d.]})]%
        {kelly_neural_2004}
\bibfield{author}{\bibinfo{person}{S. Kelly}, \bibinfo{person}{Corinne Kravitz}, {and} \bibinfo{person}{Michael Hopkins}.} \bibinfo{year}{[n.\,d.]}\natexlab{}.
\newblock \showarticletitle{Neural correlates of bimodal speech and gesture comprehension}.
\newblock   \bibinfo{volume}{89} (\bibinfo{year}{[n.\,d.]}), \bibinfo{pages}{253--260}.
\newblock
\urldef\tempurl%
\url{https://doi.org/10.1016/S0093-934X(03)00335-3}
\showDOI{\tempurl}


\bibitem[Kingma and Welling({[n.\,d.]})]%
        {kingma_auto-encoding_2022}
\bibfield{author}{\bibinfo{person}{Diederik~P. Kingma} {and} \bibinfo{person}{Max Welling}.} \bibinfo{year}{[n.\,d.]}\natexlab{}.
\newblock \bibinfo{title}{Auto-Encoding Variational Bayes}.
\newblock
\newblock
\urldef\tempurl%
\url{https://doi.org/10.48550/arXiv.1312.6114}
\showDOI{\tempurl}
\showeprint[arxiv]{1312.6114 [cs, stat]}


\bibitem[Kopp et~al\mbox{.}(2006)]%
        {koppCommonFrameworkMultimodal2006a}
\bibfield{author}{\bibinfo{person}{Stefan Kopp}, \bibinfo{person}{Brigitte Krenn}, \bibinfo{person}{Stacy Marsella}, \bibinfo{person}{Andrew~N. Marshall}, \bibinfo{person}{Catherine Pelachaud}, \bibinfo{person}{Hannes Pirker}, \bibinfo{person}{Kristinn~R. Thórisson}, {and} \bibinfo{person}{Hannes Vilhjálmsson}.} \bibinfo{year}{2006}\natexlab{}.
\newblock \showarticletitle{Towards a {Common} {Framework} for {Multimodal} {Generation}: {The} {Behavior} {Markup} {Language}}. In \bibinfo{booktitle}{\emph{Intelligent {Virtual} {Agents}}} \emph{(\bibinfo{series}{Lecture {Notes} in {Computer} {Science}})}, \bibfield{editor}{\bibinfo{person}{Jonathan Gratch}, \bibinfo{person}{Michael Young}, \bibinfo{person}{Ruth Aylett}, \bibinfo{person}{Daniel Ballin}, {and} \bibinfo{person}{Patrick Olivier}} (Eds.). \bibinfo{publisher}{Springer}, \bibinfo{address}{Berlin, Heidelberg}, \bibinfo{pages}{205--217}.
\newblock
\showISBNx{978-3-540-37594-4}
\urldef\tempurl%
\url{https://doi.org/10.1007/11821830_17}
\showDOI{\tempurl}


\bibitem[Liang et~al\mbox{.}({[n.\,d.]})]%
        {liang_seeg_2022}
\bibfield{author}{\bibinfo{person}{Yuanzhi Liang}, \bibinfo{person}{Qianyu Feng}, \bibinfo{person}{Linchao Zhu}, \bibinfo{person}{Li Hu}, \bibinfo{person}{Pan Pan}, {and} \bibinfo{person}{Yi Yang}.} \bibinfo{year}{[n.\,d.]}\natexlab{}.
\newblock \showarticletitle{{SEEG}: Semantic Energized Co-Speech Gesture Generation}. \bibinfo{pages}{10473--10482}.
\newblock
\urldef\tempurl%
\url{https://openaccess.thecvf.com/content/CVPR2022/html/Liang_SEEG_Semantic_Energized_Co-Speech_Gesture_Generation_CVPR_2022_paper.html}
\showURL{%
\tempurl}


\bibitem[Ling et~al\mbox{.}({[n.\,d.]})]%
        {ling_character_2020}
\bibfield{author}{\bibinfo{person}{Hung~Yu Ling}, \bibinfo{person}{Fabio Zinno}, \bibinfo{person}{George Cheng}, {and} \bibinfo{person}{Michiel van~de Panne}.} \bibinfo{year}{[n.\,d.]}\natexlab{}.
\newblock \showarticletitle{Character Controllers Using Motion {VAEs}}.
\newblock  \bibinfo{volume}{39}, \bibinfo{number}{4} (\bibinfo{year}{[n.\,d.]}).
\newblock
\showISSN{0730-0301, 1557-7368}
\urldef\tempurl%
\url{https://doi.org/10.1145/3386569.3392422}
\showDOI{\tempurl}
\showeprint[arxiv]{2103.14274 [cs]}


\bibitem[Liu et~al\mbox{.}({[n.\,d.]})]%
        {liu_two_2016}
\bibfield{author}{\bibinfo{person}{Kris Liu}, \bibinfo{person}{Jackson Tolins}, \bibinfo{person}{Jean~E. Fox~Tree}, \bibinfo{person}{Michael Neff}, {and} \bibinfo{person}{Marilyn~A. Walker}.} \bibinfo{year}{[n.\,d.]}\natexlab{}.
\newblock \showarticletitle{Two Techniques for Assessing Virtual Agent Personality}.
\newblock  \bibinfo{volume}{7}, \bibinfo{number}{1} (\bibinfo{year}{[n.\,d.]}), \bibinfo{pages}{94--105}.
\newblock
\showISSN{1949-3045}
\urldef\tempurl%
\url{https://doi.org/10.1109/TAFFC.2015.2435780}
\showDOI{\tempurl}


\bibitem[Lücking et~al\mbox{.}(2010)]%
        {lucking_bielefeld_2010}
\bibfield{author}{\bibinfo{person}{Andy Lücking}, \bibinfo{person}{Kirsten Bergmann}, \bibinfo{person}{Florian Hahn}, \bibinfo{person}{Stefan Kopp}, {and} \bibinfo{person}{Hannes Rieser}.} \bibinfo{year}{2010}\natexlab{}.
\newblock \showarticletitle{The Bielefeld Speech and Gesture Alignment Corpus ({SaGA})}.
\newblock  (\bibinfo{year}{2010}).
\newblock
\urldef\tempurl%
\url{https://pub.uni-bielefeld.de/record/2001935}
\showURL{%
\tempurl}


\bibitem[Lücking et~al\mbox{.}(2013)]%
        {lucking_data-based_2013}
\bibfield{author}{\bibinfo{person}{Andy Lücking}, \bibinfo{person}{Kirsten Bergmann}, \bibinfo{person}{Florian Hahn}, \bibinfo{person}{Stefan Kopp}, {and} \bibinfo{person}{Hannes Rieser}.} \bibinfo{year}{2013}\natexlab{}.
\newblock \showarticletitle{Data-based analysis of speech and gesture: the Bielefeld Speech and Gesture Alignment corpus ({SaGA}) and its applications}.
\newblock  \bibinfo{volume}{7}, \bibinfo{number}{1} (\bibinfo{year}{2013}).
\newblock
\showISSN{1783-7677}
\urldef\tempurl%
\url{https://pub.uni-bielefeld.de/record/2522299}
\showURL{%
\tempurl}


\bibitem[Macedonia({[n.\,d.]})]%
        {macedonia_imitation_2014}
\bibfield{author}{\bibinfo{person}{Manuela Macedonia}.} \bibinfo{year}{[n.\,d.]}\natexlab{}.
\newblock \showarticletitle{Imitation of a Pedagogical Agent’s Gestures Enhances Memory for Words in Second Language}.
\newblock  \bibinfo{volume}{2}, \bibinfo{number}{5} (\bibinfo{year}{[n.\,d.]}), \bibinfo{pages}{162}.
\newblock
\showISSN{2329-0900}
\urldef\tempurl%
\url{https://doi.org/10.11648/j.sjedu.20140205.15}
\showDOI{\tempurl}


\bibitem[McNeill et~al\mbox{.}(1990)]%
        {mcneill1990speech}
\bibfield{author}{\bibinfo{person}{David McNeill}, \bibinfo{person}{Laura~L Pedelty}, {and} \bibinfo{person}{Elena~T Levy}.} \bibinfo{year}{1990}\natexlab{}.
\newblock \showarticletitle{Speech and gesture}.
\newblock In \bibinfo{booktitle}{\emph{Advances in psychology}}. Vol.~\bibinfo{volume}{70}. \bibinfo{publisher}{Elsevier}, \bibinfo{pages}{203--256}.
\newblock


\bibitem[Neff et~al\mbox{.}({[n.\,d.]})]%
        {neff_evaluating_2010}
\bibfield{author}{\bibinfo{person}{Michael Neff}, \bibinfo{person}{Yingying Wang}, \bibinfo{person}{Rob Abbott}, {and} \bibinfo{person}{Marilyn Walker}.} \bibinfo{year}{[n.\,d.]}\natexlab{}.
\newblock \bibinfo{booktitle}{\emph{Evaluating the Effect of Gesture and Language on Personality Perception in Conversational Agents}}.
\newblock
\showISBNx{978-3-642-15891-9}
\urldef\tempurl%
\url{https://doi.org/10.1007/978-3-642-15892-6_24}
\showDOI{\tempurl}
\newblock
\shownote{Pages: 235}.


\bibitem[Nyatsanga et~al\mbox{.}({[n.\,d.]})]%
        {nyatsanga_comprehensive_2023}
\bibfield{author}{\bibinfo{person}{Simbarashe Nyatsanga}, \bibinfo{person}{Taras Kucherenko}, \bibinfo{person}{Chaitanya Ahuja}, \bibinfo{person}{Gustav~Eje Henter}, {and} \bibinfo{person}{Michael Neff}.} \bibinfo{year}{[n.\,d.]}\natexlab{}.
\newblock \showarticletitle{A Comprehensive Review of Data-Driven Co-Speech Gesture Generation}.
\newblock  \bibinfo{volume}{42}, \bibinfo{number}{2} (\bibinfo{year}{[n.\,d.]}), \bibinfo{pages}{569--596}.
\newblock
\showISSN{0167-7055, 1467-8659}
\urldef\tempurl%
\url{https://doi.org/10.1111/cgf.14776}
\showDOI{\tempurl}
\showeprint[arxiv]{2301.05339 [cs]}


\bibitem[Oord et~al\mbox{.}({[n.\,d.]})]%
        {oord_neural_2018}
\bibfield{author}{\bibinfo{person}{Aaron van~den Oord}, \bibinfo{person}{Oriol Vinyals}, {and} \bibinfo{person}{Koray Kavukcuoglu}.} \bibinfo{year}{[n.\,d.]}\natexlab{}.
\newblock \bibinfo{title}{Neural Discrete Representation Learning}.
\newblock
\newblock
\urldef\tempurl%
\url{https://doi.org/10.48550/arXiv.1711.00937}
\showDOI{\tempurl}
\showeprint[arxiv]{1711.00937 [cs]}


\bibitem[Pandey and Gelin({[n.\,d.]})]%
        {pandey_mass-produced_2018}
\bibfield{author}{\bibinfo{person}{Amit~Kumar Pandey} {and} \bibinfo{person}{Rodolphe Gelin}.} \bibinfo{year}{[n.\,d.]}\natexlab{}.
\newblock \showarticletitle{A Mass-Produced Sociable Humanoid Robot: Pepper: The First Machine of Its Kind}.
\newblock  \bibinfo{volume}{25}, \bibinfo{number}{3} (\bibinfo{year}{[n.\,d.]}), \bibinfo{pages}{40--48}.
\newblock
\showISSN{1558-223X}
\urldef\tempurl%
\url{https://doi.org/10.1109/MRA.2018.2833157}
\showDOI{\tempurl}
\newblock
\shownote{Conference Name: {IEEE} Robotics \& Automation Magazine}.


\bibitem[Press(2024a)]%
        {noauthor_direction_nodate}
\bibfield{author}{\bibinfo{person}{Oxford~University Press}.} \bibinfo{year}{2024}\natexlab{a}.
\newblock \bibinfo{booktitle}{\emph{direction, n. meanings, etymology and more {\textbar} Oxford English Dictionary}}.
\newblock
\urldef\tempurl%
\url{https://www.oed.com/dictionary/direction_n}
\showURL{%
\tempurl}


\bibitem[Press(2024b)]%
        {noauthor_movement_nodate}
\bibfield{author}{\bibinfo{person}{Oxford~University Press}.} \bibinfo{year}{2024}\natexlab{b}.
\newblock \bibinfo{booktitle}{\emph{movement, n. meanings, etymology and more {\textbar} Oxford English Dictionary}}.
\newblock
\urldef\tempurl%
\url{https://www.oed.com/dictionary/movement_n}
\showURL{%
\tempurl}


\bibitem[Press(2024c)]%
        {noauthor_negation_nodate}
\bibfield{author}{\bibinfo{person}{Oxford~University Press}.} \bibinfo{year}{2024}\natexlab{c}.
\newblock \bibinfo{booktitle}{\emph{negation, n. meanings, etymology and more {\textbar} Oxford English Dictionary}}.
\newblock
\urldef\tempurl%
\url{https://www.oed.com/dictionary/negation_n}
\showURL{%
\tempurl}


\bibitem[Press(2024d)]%
        {noauthor_object_nodate}
\bibfield{author}{\bibinfo{person}{Oxford~University Press}.} \bibinfo{year}{2024}\natexlab{d}.
\newblock \bibinfo{booktitle}{\emph{object, n. meanings, etymology and more {\textbar} Oxford English Dictionary}}.
\newblock
\urldef\tempurl%
\url{https://www.oed.com/dictionary/object_n}
\showURL{%
\tempurl}


\bibitem[Press(2024e)]%
        {noauthor_shape_nodate}
\bibfield{author}{\bibinfo{person}{Oxford~University Press}.} \bibinfo{year}{2024}\natexlab{e}.
\newblock \bibinfo{booktitle}{\emph{shape, n.¹ meanings, etymology and more {\textbar} Oxford English Dictionary}}.
\newblock
\urldef\tempurl%
\url{https://www.oed.com/dictionary/shape_n1}
\showURL{%
\tempurl}


\bibitem[Press(2024f)]%
        {noauthor_size_nodate}
\bibfield{author}{\bibinfo{person}{Oxford~University Press}.} \bibinfo{year}{2024}\natexlab{f}.
\newblock \bibinfo{booktitle}{\emph{size, n.¹ meanings, etymology and more {\textbar} Oxford English Dictionary}}.
\newblock
\urldef\tempurl%
\url{https://www.oed.com/dictionary/size_n1}
\showURL{%
\tempurl}


\bibitem[Razavi et~al\mbox{.}({[n.\,d.]})]%
        {razavi_generating_2019}
\bibfield{author}{\bibinfo{person}{Ali Razavi}, \bibinfo{person}{Aaron van~den Oord}, {and} \bibinfo{person}{Oriol Vinyals}.} \bibinfo{year}{[n.\,d.]}\natexlab{}.
\newblock \bibinfo{title}{Generating Diverse High-Fidelity Images with {VQ}-{VAE}-2}.
\newblock
\newblock
\urldef\tempurl%
\url{https://doi.org/10.48550/arXiv.1906.00446}
\showDOI{\tempurl}
\showeprint[arxiv]{1906.00446 [cs, stat]}


\bibitem[Salem et~al\mbox{.}({[n.\,d.]})]%
        {salem_generation_2012}
\bibfield{author}{\bibinfo{person}{Maha Salem}, \bibinfo{person}{Stefan Kopp}, \bibinfo{person}{Ipke Wachsmuth}, \bibinfo{person}{Katharina Rohlfing}, {and} \bibinfo{person}{Frank Joublin}.} \bibinfo{year}{[n.\,d.]}\natexlab{}.
\newblock \showarticletitle{Generation and Evaluation of Communicative Robot Gesture}.
\newblock  \bibinfo{volume}{4}, \bibinfo{number}{2} (\bibinfo{year}{[n.\,d.]}), \bibinfo{pages}{201--217}.
\newblock
\showISSN{1875-4791, 1875-4805}
\urldef\tempurl%
\url{https://doi.org/10.1007/s12369-011-0124-9}
\showDOI{\tempurl}


\bibitem[Vaswani et~al\mbox{.}({[n.\,d.]})]%
        {vaswani_attention_2023}
\bibfield{author}{\bibinfo{person}{Ashish Vaswani}, \bibinfo{person}{Noam Shazeer}, \bibinfo{person}{Niki Parmar}, \bibinfo{person}{Jakob Uszkoreit}, \bibinfo{person}{Llion Jones}, \bibinfo{person}{Aidan~N. Gomez}, \bibinfo{person}{Lukasz Kaiser}, {and} \bibinfo{person}{Illia Polosukhin}.} \bibinfo{year}{[n.\,d.]}\natexlab{}.
\newblock \bibinfo{title}{Attention Is All You Need}.
\newblock
\newblock
\urldef\tempurl%
\url{https://doi.org/10.48550/arXiv.1706.03762}
\showDOI{\tempurl}
\showeprint[arxiv]{1706.03762 [cs]}


\bibitem[Vo{\ss} and Kopp(2023)]%
        {voss2023aq}
\bibfield{author}{\bibinfo{person}{Hendric Vo{\ss}} {and} \bibinfo{person}{Stefan Kopp}.} \bibinfo{year}{2023}\natexlab{}.
\newblock \showarticletitle{AQ-GT: a temporally aligned and quantized GRU-transformer for co-speech gesture synthesis}. In \bibinfo{booktitle}{\emph{Proceedings of the 25th International Conference on Multimodal Interaction}}. \bibinfo{pages}{60--69}.
\newblock


\bibitem[Voß and Kopp(2023)]%
        {vos_augmented_2023}
\bibfield{author}{\bibinfo{person}{Hendric Voß} {and} \bibinfo{person}{Stefan Kopp}.} \bibinfo{year}{2023}\natexlab{}.
\newblock \bibinfo{title}{Augmented Co-Speech Gesture Generation: Including Form and Meaning Features to Guide Learning-Based Gesture Synthesis}.
\newblock
\newblock
\urldef\tempurl%
\url{https://doi.org/10.48550/arXiv.2307.09597}
\showDOI{\tempurl}
\showeprint[arxiv]{2307.09597 [cs]}


\bibitem[Wagner et~al\mbox{.}({[n.\,d.]})]%
        {wagner_gesture_2014}
\bibfield{author}{\bibinfo{person}{Petra Wagner}, \bibinfo{person}{Zofia Malisz}, {and} \bibinfo{person}{Stefan Kopp}.} \bibinfo{year}{[n.\,d.]}\natexlab{}.
\newblock \showarticletitle{Gesture and speech in interaction: An overview}.
\newblock   \bibinfo{volume}{57} (\bibinfo{year}{[n.\,d.]}), \bibinfo{pages}{209--232}.
\newblock
\showISSN{01676393}
\urldef\tempurl%
\url{https://doi.org/10.1016/j.specom.2013.09.008}
\showDOI{\tempurl}


\bibitem[Wu et~al\mbox{.}({[n.\,d.]})]%
        {wu_wasserstein_2018}
\bibfield{author}{\bibinfo{person}{Jiqing Wu}, \bibinfo{person}{Zhiwu Huang}, \bibinfo{person}{Janine Thoma}, \bibinfo{person}{Dinesh Acharya}, {and} \bibinfo{person}{Luc Van~Gool}.} \bibinfo{year}{[n.\,d.]}\natexlab{}.
\newblock \bibinfo{title}{Wasserstein Divergence for {GANs}}.
\newblock
\newblock
\urldef\tempurl%
\url{https://doi.org/10.48550/arXiv.1712.01026}
\showDOI{\tempurl}
\showeprint[arxiv]{1712.01026 [cs]}


\bibitem[Yoon et~al\mbox{.}(2020)]%
        {yoon_speech_2020}
\bibfield{author}{\bibinfo{person}{Youngwoo Yoon}, \bibinfo{person}{Bok Cha}, \bibinfo{person}{Joo-Haeng Lee}, \bibinfo{person}{Minsu Jang}, \bibinfo{person}{Jaeyeon Lee}, \bibinfo{person}{Jaehong Kim}, {and} \bibinfo{person}{Geehyuk Lee}.} \bibinfo{year}{2020}\natexlab{}.
\newblock \showarticletitle{Speech Gesture Generation from the Trimodal Context of Text, Audio, and Speaker Identity}.
\newblock  \bibinfo{volume}{39}, \bibinfo{number}{6} (\bibinfo{year}{2020}), \bibinfo{pages}{1--16}.
\newblock
\showISSN{0730-0301, 1557-7368}
\urldef\tempurl%
\url{https://doi.org/10.1145/3414685.3417838}
\showDOI{\tempurl}
\showeprint[arxiv]{2009.02119 [cs]}


\bibitem[Yoon et~al\mbox{.}({[n.\,d.]})]%
        {yoon_sgtoolkit_2021}
\bibfield{author}{\bibinfo{person}{Youngwoo Yoon}, \bibinfo{person}{Keunwoo Park}, \bibinfo{person}{Minsu Jang}, \bibinfo{person}{Jaehong Kim}, {and} \bibinfo{person}{Geehyuk Lee}.} \bibinfo{year}{[n.\,d.]}\natexlab{}.
\newblock \showarticletitle{{SGToolkit}: An Interactive Gesture Authoring Toolkit for Embodied Conversational Agents}. In \bibinfo{booktitle}{\emph{The 34th Annual {ACM} Symposium on User Interface Software and Technology}} (2021-10-10). \bibinfo{pages}{826--840}.
\newblock
\urldef\tempurl%
\url{https://doi.org/10.1145/3472749.3474789}
\showDOI{\tempurl}
\showeprint[arxiv]{2108.04636 [cs]}


\end{thebibliography}

\end{document}